\documentclass[citeautoscript,floatfix,superscriptaddress,twocolumn,showpacs,prb,aps,10pt]{revtex4-1}

\usepackage{graphicx} \usepackage{amsmath, amssymb, bm}
\usepackage{dcolumn} \usepackage{color}
\usepackage{eso-pic}
\usepackage{fix-cm}

\usepackage{multirow}
\usepackage{sidecap}
\usepackage{marginnote}
\usepackage{ulem} 

\newcommand{\sss}{\scriptscriptstyle}
\newcommand{\sst}{\scriptstyle}

\newcommand{\stext}[1]{\sss \text{#1} \sst}

\renewcommand{\emph}[1]{\textit{#1}}

\begin{document}
\title{Infrared active phonons in monoclinic lutetium oxyorthosilicate}

\author{M. Stokey}
\email{mstokey@huskers.unl.edu}
\homepage{http://ellipsometry.unl.edu}
\affiliation{Department of Electrical and Computer Engineering, University of Nebraska-Lincoln, Lincoln, NE 68588, USA}
\author{A. Mock}
\affiliation{National Research Council Postdoctoral Fellow, residing at U.S. Naval Research Laboratory, Washington, DC 20375, USA}
\author{R. Korlacki}
\affiliation{Department of Electrical and Computer Engineering, University of Nebraska-Lincoln, Lincoln, NE 68588, USA}
\author{S. Knight}
\affiliation{Department of Electrical and Computer Engineering, University of Nebraska-Lincoln, Lincoln, NE 68588, USA}
\author{V. Darakchieva}
\affiliation{Terahertz Materials Analysis Center and Competence Center for III-Nitride Technology C3NiT - Janz\'{e}n, Department of Physics, Chemistry and Biology (IFM), Link\"{o}ping University, Link\"{o}ping, SE 58183, Sweden}
\author{S. Sch\"{o}che}
\affiliation{J.A. Woollam Co. Lincoln, NE 68508, USA}
\author{M. Schubert}
\affiliation{Department of Electrical and Computer Engineering, University of Nebraska-Lincoln, Lincoln, NE 68588, USA}
\affiliation{Terahertz Materials Analysis Center and Competence Center for III-Nitride Technology C3NiT - Janz\'{e}n, Department of Physics, Chemistry and Biology (IFM), Link\"{o}ping University, Link\"{o}ping, SE 58183, Sweden}
\affiliation{Leibniz Institute for Polymer Research, Dresden, D 01069, Germany}
\date{\today}

\begin{abstract}
A combined generalized spectroscopic ellipsometry measurement and density functional theory calculation analysis is performed to obtain the complete set of infrared active phonon modes in Lu$_{2}$SiO$_{5}$ with monoclinic crystal structure. Two different crystals, each cut perpendicular to a different crystal axis are investigated. Ellipsometry measurements from 40 - 1200 cm$^{-1}$ are used to determine the frequency dependent dielectric function tensor elements. The eigendielectric displacement vector summation approach and eigendielectric displacement loss vector summation approach both augmented with anharmonic lattice broadening parameters proposed recently for low-symmetry crystal structures [A. Mock~\textit{et al.}, Phys. Rev. B 95, 165202 (2017)] are applied for our ellipsometry data analysis. All measured and model calculated dielectric function tensor and inverse dielectric function tensor elements match excellently. 23 $\mathrm{A_{u}}$ symmetry and 22 $\mathrm{B_{u}}$ symmetry infrared active transverse and longitudinal optical modes are found. We also determine the directional limiting modes and the order of the phonon modes within the monoclinic plane. Results from density functional theory and ellipsometry measurements are compared and nearly perfect agreement is observed. We further compare our results to those obtained recently for the monoclinic crystal Y$_{2}$SiO$_{5}$, which is isostructural to Lu$_{2}$SiO$_{5}$ [A. Mock~\textit{et al.}, Phys. Rev. B 97, 165203 (2018)]. We find that the lattice mode behavior of monoclinic Lu$_{2}$SiO$_{5}$ is qualitatively identical with Y$_{2}$SiO$_{5}$, and differs only quantitatively. We anticipate that members of the isostructural group of monoclinic symmetry oxyorthosilicates such as Dy$_{2}$SiO$_{5}$ or Yb$_{2}$SiO$_{5}$ will likely behave very similar in their phonon mode properties as reported here for Lu$_{2}$SiO$_{5}$.
\end{abstract}

\maketitle

\section {Introduction}
Rare-earth oxyorthosilicates offer a unique set of characteristics that make them well suited to a broad range of applications. These oxyorthosilicates include Dy$_{2}$SiO$_{5}$, Lu$_{2}$SiO$_{5}$ (LSO), Yb$_{2}$SiO$_{5}$, among others.\cite{Felsche_1973} These materials are isostructural also with Y$_{2}$SiO$_{5}$ (YSO) and both similarities as well as differences are anticipated in their optical properties. LSO has been of interest as a scintillator material when doped with cerium.\cite{Melcher92} Oxyorthosilicates such as LSO are attractive candidates for scintillators due to their large band gap energy value and large mass density. To withstand high energy gamma radiation, scintillators should have a density over 7~gm/cm$^{3}$.\cite{LEMPICKI1998} A fast decay time below 100~ns, and a light output over 8500~photons/MeV are required to meet demands for fast and efficient detection.\cite{LEMPICKI1998} Undoped LSO is an insulator with a band gap energy near 6~eV. \cite{Naud_1996} Dopant states within the band gap serve as activation centers to produce photons exploiting, for example, $4f-5d$ transitions.\cite{Zavartsev_2005} In Ce$^{3+}$ doped LSO, where Ce replaces Lu, the 4f$^{1}$ band is split by spin-orbit interactions into levels $^{2}$F$_{5/2}$ and $^{2}$F$_{7/2}$ leading to transition wavelengths of 360~nm, 300~nm, and 260~nm. \cite{GRYK2006} Very little is known presently about fundamental physics properties of LSO in particular and oxyorthosilicates in general.

\begin{figure}[hbt]
\centering
\includegraphics[width=0.5\textwidth]{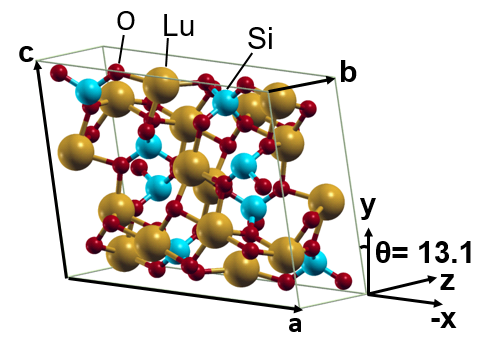}
\caption{\label{fig:unitcell}(a) Unit cell of Lu$_2$SiO$_5$, with monoclinic axes ($a$,$b$,$c$). Overlaid is the Cartesian sample system ($x$,$y$,$z$). The laboratory Cartesian coordinate system ($\hat{x}$,$\hat{y}$,$\hat{z}$) (not shown here) is associated with the ellipsometer instrument, where a given sample surface is parallel to plane $\hat{x}$ - $\hat{y}$ and at $\hat{z}$=0, the plane of incidence is parallel $\hat{x}$.} 
\end{figure}

LSO and the isostructural oxyorthosilicates crystallize as  monoclinic crystal in space group 15, with 18 possible choices to address the unit cell. When comparing results from different reports, care must be taken to properly transform unit axes descriptions. Density functional theory (DFT) calculations have been used to predict lattice parameters of oxyorthosilicates,\cite{YLi_2018, Zhang_2015, Ning2012, TIAN2015} and comparison with experiment was made available by X-ray diffraction (XRD) analyses of powdered samples.\cite{Glowacki_2012, Chiriu_2007, Dominiak_2010,  KITAURA2015, TIAN2015, GRYK2006, Cong2009, Gustafsson2001} Key to further understanding is the availability of high quality crystals. Growth using the Czochralski method\cite{Melcher92} results in single crystals and deposition using high-pressure annealing techniques results in ceramics.\cite{Roy_2013} While single crystal growth is slow and energy extensive, ceramics show inferior optical properties.\cite{Roy_2013,TIAN2015} Initial 
Czochralski growth used iridium wire, while the availability of seed crystals improved quality. Presently, high quality single crystals with various doping are available commercially, while some laboratories have grown their own material.\cite{Jellison_2012,Voronko_2011, Cong2009}

LSO has gained substantial attraction in nuclear medicine for use as scintillator material in positron emission tomography and combined positron emission tomography x-ray computed-tomography.\cite{Melcher_2000,Michail_2008, Chatziioannou_1999,Pepin2004,SATO2011} A large gamma radiation to photon conversion efficiency and a very short decay time offer increased spatial and temporal resolution in medical imaging also permitting shorter image acquisition time.\cite{Melcher92,Melcher_2000} Its high efficiency also permits development of smaller devices.\cite{Chatziioannou_1999} 
Likewise, LSO has potential for use in high-energy physics. Chen~\textit{et al.} demonstrated that continued exposure to gamma rays did not permanently alter its optical properties.\cite{Chen_2007} Radiation hardness is required for use in calorimeters, which measure the energy of particles produced by collisions in accelerators. LSO was proposed for the now cancelled SuperB collider project, the Mu2e collider, and in an upgrade to the Super Large Hadron Collider.\cite{Eigen2013, Pezzullo_2014, Zhu2017}
LSO is also currently investigated as active laser material. While Ce:LSO is not a commonly used media yet, similar lanthanide doped orthosilicates have been used as effective high-powered lasers. For example, Yb:YSO, and Yb:(Lu$_x$Y$_{(1-x)}$)SO have been shown to be highly efficient continuous-wave laser materials with emission wavelengths near or above 1000~nm.\cite{Li_2006,WLi_2008,Brickeen2009} Yb:LSO has a remarkably high saturation density of 9.2~kW/cm$^{2}$ and an emission wavelength of 1083~nm.\cite{Zheng_2009} Neodymium doped LSO is a promising laser material with a unique emission wavelength of 1357~nm.\cite{Zhuang2012, Cong_2010} Nd:LSO has been proposed as a pump for strontium optical clocks, laser doppler velocimetry, and distributed fiber sensor applications.\cite{Zhuang2012} Yb:LSO has been proposed for femtosecond pulsed laser applications by Thibault~\textit{et al.}, who report optical pump pulse efficiency of 17$\%$. \cite{Thibault2006}

Monoclinic crystals are highly anisotropic and must be described by four independent dielectric function tensor element spectra, all of which contain real and imaginary parts. Spectra of the optical properties of monoclinic crystals have only been reported very recently, and generalized spectroscopic ellipsometry techniques for proper measurements and data analysis of arbitrarily anisotropic materials have only been developed recently.\cite{SchubertADP15_2006} For example, monoclinic $\beta$-Ga$_2$O$_3$, which has gained attention for high power solid state transistor and switch applications, has been analyzed from the terahertz to the vacuum ultra violet spectral regions for its complete phonon mode and free charge carrier properties,\cite{Schubert2016,KnightAPLbGOOHE2018,SchubertAPLLPPbGO2019,SchubertPRB2019} and electronic band-to-band transitions as well as its static and high-frequency dielectric constants, indices of refraction, and extinction coefficients.\cite{Mock_2017Ga2O3,Sturm_2015,Sturm_2016,Sturm_2017,MockAPLbGO2019} A similar analysis procedure resulted in complete sets of phonon modes for monoclinic scintillator material cadmium tungstate (CdWO$_4$)\cite{Mock_2016} and YSO.\cite{Mock2018} To date, the only generalized spectroscopic ellipsometry investigation of LSO was reported by Jellison~\textit{et al.}, and the four real-valued dielectric function tensor spectra were obtained in the spectral range from 200~nm to 850~nm.\cite{Jellison_2012} The spectral dependencies of the monoclinic refractive indices were calculated and the band gap of LSO was estimated to be at 6.7~eV. Kitaura~\textit{et al.} reported vacuum ultra violet reflectivity spectra to nearly 30~eV\cite{KITAURA2015} and provided an in-depth study of the electronic band-to-band transitions in LSO. Kitaura~\textit{et al.} estimated the band gap to be at 7.52~eV. 

Lattice mode excitations are fundamental physical processes which influence many material properties such as thermal and electronic transport, or coupling with excitons or photon emission. The lattice thermal conductivity of LSO is important when used as active laser material, or as an environmental barrier coating.\cite{Cong2009, YLi_2018} Since LSO is highly anisotropic, the thermal conductivity should also depend on transport direction. Gustafsson~\textit{et al.} reported on the thermal properties of LSO but assumed isotropic behavior.\cite{Gustafsson2001} Subsequently, Cong~\textit{et al.} showed the effects of anisotropy in thermal conductivity and reported the behavior along each crystal axis.\cite{Cong2009} While optical phonons do not directly participate in heat transfer, it is theorized that the anisotropic transport originates from coupling with anisotropic acoustic phonons.\cite{Lou_2016} Raman active modes have been well studied for a large range of oxyorthosilicates including LSO.\cite{Glowacki_2012, Ricci_Press_2008, Ricci_2008, Chiriu_2007, Binczyk2016, Voronko_2011} However, in many of these reports the crystals are approximated as isotropic materials ignoring wavevector and polarization dependencies of the Raman modes. Infrared active modes were not  reported for LSO. Likewise, no DFT calculations of lattice mode properties have been reported.

In this work we report a combined generalized spectroscopic ellipsometry and DFT analysis to obtain the complete set of long wavelength active phonon modes in cerium doped Lu$_{2}$SiO$_{5}$. Ellipsometry measurements from 40 - 1200 cm$^{-1}$ are used to determine the frequency dependent dielectric function tensor elements. Two different crystals, each with different cuts perpendicular to a crystal axis are investigated. We closely follow the methodology recently described for YSO by Mock~\textit{et al.} in order to determine the complete set of long wavelength active phonon modes.\cite{Mock2018} We expect qualitatively similar results for LSO since YSO is isostructural. As such, identical numbers of phonon pairs are expected. However, both materials differ quantitatively, which will be discussed in more detail. We find and detail 23 A$_\mathrm{u}$ symmetry and 22 B$_\mathrm{u}$ symmetry infrared active transverse (TO) and longitudinal optical (LO) modes. We also perform DFT calculations and compare our results with the phonon modes from experiment. An excellent agreement is obtained. We also compare the phonon mode frequencies and the phonon mode order within LSO and YSO and find strong qualitative agreement. We therefore anticipate similar phonon mode properties among members of the isostructural group of monoclinic symmetry rare-earth oxyorthosilicates such as Dy$_{2}$SiO$_{5}$, Ho$_{2}$SiO$_{5}$, Er$_{2}$SiO$_{5}$, Tm$_{2}$SiO$_{5}$, and Yb$_{2}$SiO$_{5}$.\cite{Felsche_1973} We briefly describe the eigendielectric vector summation approaches for rendering the infrared optical properties of monoclinic crystals including lattice anharmonicity and the generalized spectroscopic ellipsometry measurements. We detail our DFT approach, report all results, and discuss our findings. In our ellipsometry analysis we did not observe additional modes due to the presence of the dopant cerium. Hence, throughout this work, the effects of the dopant cerium onto the lattice modes are ignored. The results of our work will become relevant for the future understanding of optical properties such as photon-phonon coupling and lattice thermal transport processes.

\section{Theory}
\subsection{Structure and Symmetry} 
Lu$_{2}$SiO$_{5}$ belongs to the space group 15 (centered monoclinic). There are 18 alternative choices for the unit cell for this space group, per the International Tables for Crystallography. \cite{ITA92} The crystallographic standard for the unit cell choice for monoclinic crystals\cite{Kennard_1967,mighell02} promotes choosing vector \textbf{b} along the symmetry axis, and the lowest possible non-acute monoclinic angle in the network perpendicular to the symmetry axis. For space group 15 the $I2/c$ cell is consistent with the standard and is used throughout this work. The structural parameters for the unit cell are specified in the next section. In many previous publications, the $C2/c$ cell was chosen.\cite{Melcher92,Gustafsson2001,Voronko_2011,Ning2012,TIAN2015,Jellison_2012} Where appropriate, we convert the literature unit cell parameters to the $I2/c$ cell used here. The $I2/c$ cell definition is shown in Fig.~\ref{fig:unitcell}.

Four lutetium atoms form a distorted tetrahedron with isolated ionic units of SiO$_4$ tetrahedrals and oxygen atoms not bonded to silicon (Fig.~\ref{fig:unitcell}).\cite{Gustafsson2001} The crystallographic axis c is distinguished by OLu$_4$ tetrahedra along edge-sharing infinite chains. Two crystallographic sites, Lu$_1$ and Lu$_2$, coordinated with either six or seven oxygen atoms, respectively, are occupied by Lu$^{3+}$ ions.

\subsection{Density Functional Theory}
\begin{figure*}[hbt]
\centering
\includegraphics[width=1.0\linewidth]{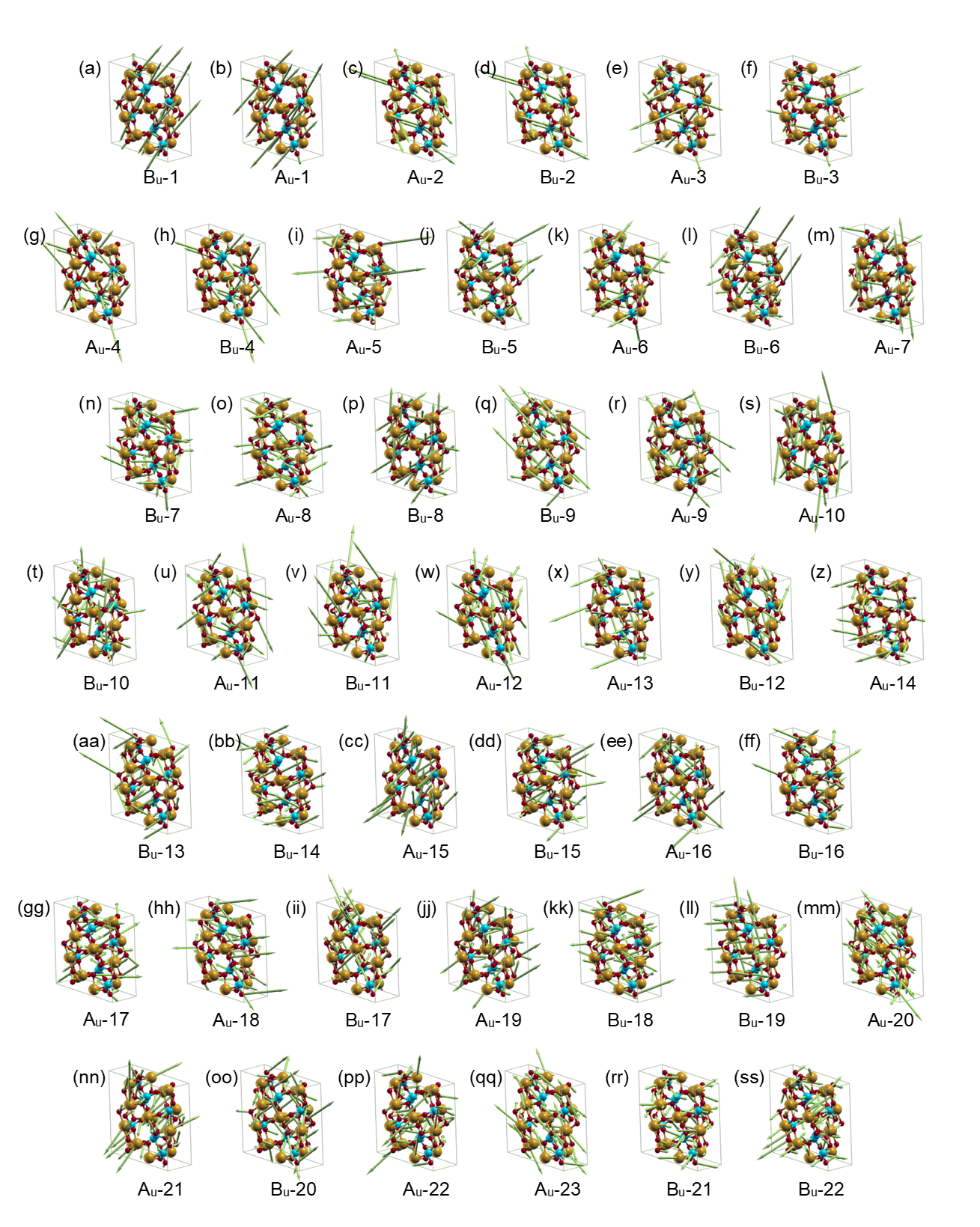}
\caption{\label{fig:TOphonons} DFT calculated phonon mode displacements for TO modes identified in this work for Lu$_2$SiO$_5$. Shown here are modes with $\mathrm{B_u}$ (See also Tab.~\ref{tab:BuDFT}) and $\mathrm{A_u}$ (Tab.~\ref{tab:AuDFT}) symmetry. The order of the phonon mode labelling is performed with increasing wavelength.}
\end{figure*}
\begin{figure*}[hbt]
\centering
\includegraphics[width=1.0\linewidth]{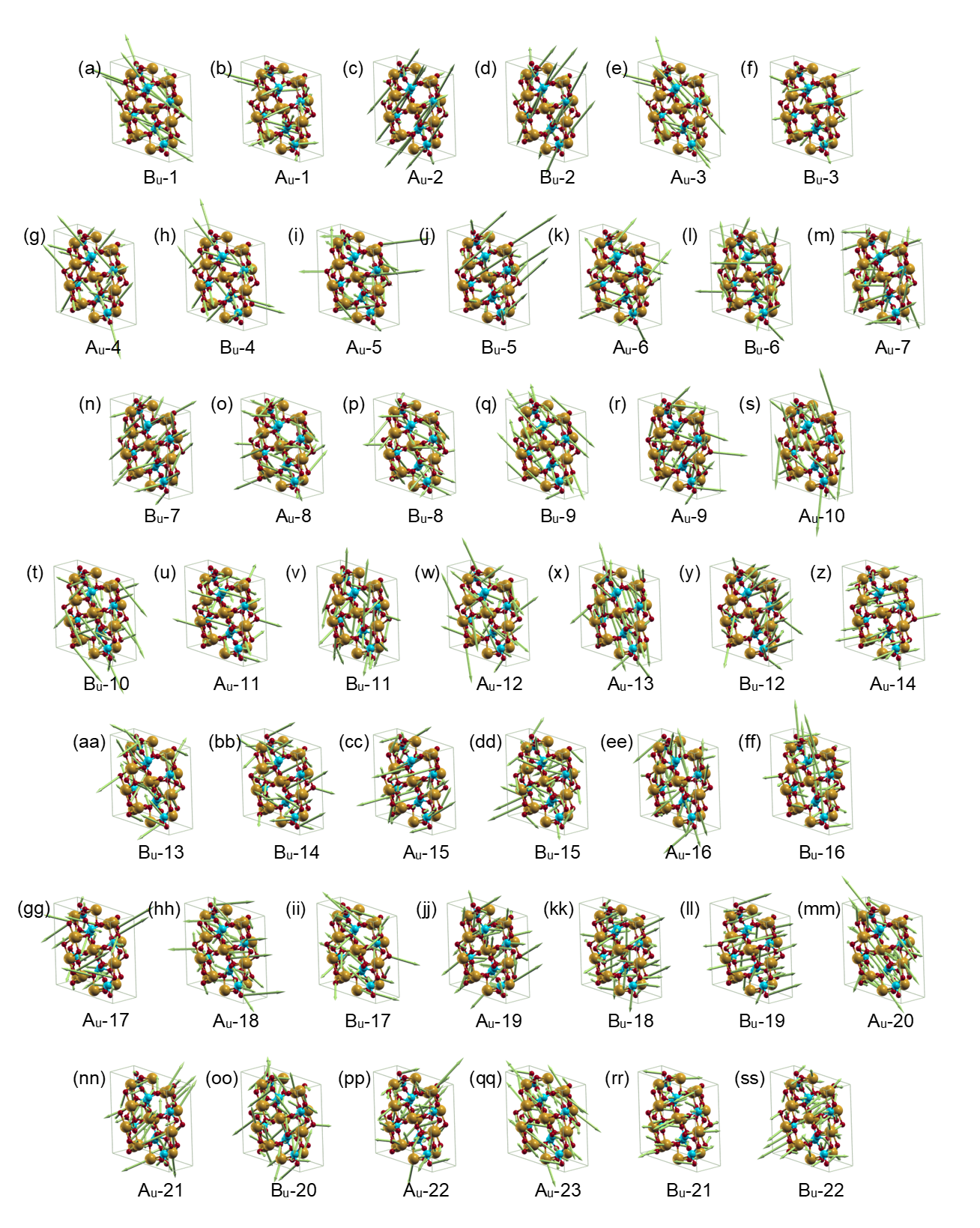}
\caption{\label{fig:LOphonons}  Same as Fig.~\ref{fig:TOphonons} for LO modes.} 
\end{figure*}

\begin{table}
\caption{\label{tab:lattice} Comparison between the experimental and theoretical lattice constants (in \AA; monoclinic angle $\beta$ in $^{\circ}$).}
\begin{ruledtabular}
\begin{tabular}{lccccccc}
&Exp.$^\textrm{a}$&Exp.$^\textrm{b}$&Calc.$^\textrm{c}$&Calc.$^\textrm{c}$&Calc.$^\textrm{e}$\\
\hline
\textbf{a}& 12.362 & 12.363 & 12.36 & 12.409 & 12.385 \\
\textbf{b} & 6.640 & 6.647 & 6.63 & 6.669 & 6.650 \\
\textbf{c} & 10.247 & 10.255 & 10.27 & 10.284 & 10.273\\
$\beta$ & 102.299 & 102.422 & 101.84 & 102.622 & 103.119 \\
\end{tabular}
\end{ruledtabular}
\begin{flushleft}
\footnotesize{$^\textrm{a}${Ref.~\onlinecite{Gustafsson2001}.}}\\
\footnotesize{$^\textrm{b}${Ref.~\onlinecite{Cong2009}.}}\\
\footnotesize{$^\textrm{c}${Ref.~\onlinecite{TIAN2015}, LDA.}}\\
\footnotesize{$^\textrm{d}${Ref.~\onlinecite{Ning2012}, GGA-PBE.}}\\
\footnotesize{$^\textrm{e}${This work, GGA-PBE.}}\\
\end{flushleft}
\end{table}

Theoretical calculations were performed by plane wave DFT code Quantum ESPRESSO (QE).\cite{[{Quantum ESPRESSO is available from http://www.qu\-an\-tum-es\-pres\-so.org. See also: }]GiannozziJPCM2009QE} We used the exchange correlation functional of Perdew, Burke, and Ernzerhof (PBE)\cite{PBE}, and optimized norm-conserving Vanderbilt (ONCV) scalar-relativistic pseudopotentials.\cite{Hamann2013,vanSetten2018} For lutetium we used a pseudopotential with the \textit{f} states frozen into the core. As the Lu$_2$SiO$_5$ is isostructural with many similar rare earth oxyorthosilicates\cite{Felsche_1973} as well as with the Y$_2$SiO$_5$, we used the optimized crystal structure of the latter from our recent study\cite{Mock2018} as the starting point for the calculations of Lu$_2$SiO$_5$. The calculations were performed in a primitive cell \textbf{p$_1$} = \textbf{a}, \textbf{p$_2$} = \textbf{b}, \textbf{p$_3$} =(\textbf{a+b+c})/2. The initial structure was first relaxed to force levels less than 10$^{-5}$ Ry Bohr$^{-1}$. A regular shifted $4\times4\times4$ Monkhorst-Pack grid was used for sampling of the Brillouin zone.\cite{MonkhorstPRBGRID} A convergence threshold of $1\times10^{-12}$  Ry was used to reach self consistency with a large electronic wavefunction cutoff of 120 Ry. The comparison of resulting optimized cell parameters with the existing literature data are listed in Table \ref{tab:lattice}. The relaxed cell was used for subsequent phonon calculations, which are described in Section \ref{dft_phonons}. 

\subsection{Infrared dielectric function tensor models}

We use the eigendielectric displacement vector summation (EDVS)\cite{Schubert2016,Schubert_2016_LST,Mock_2017} approach and eigendielectric displacement loss vector summation (EDLVS) approach augmented with anharmonic lattice broadening parameters for analysis of the dielectric function tensor data extracted from the ellipsometry measurements. The process for extracting the tensor data from the experiment is discussed further below. The EDVS approach permits access to the TO mode parameters. The EDVLS approach permits access to the LO mode parameters. The application of both approaches simultaneously in order to determine the amplitudes, frequencies and broadening parameters of all TO and LO modes was shown previously by Mock~\textit{et al.} for YSO,\cite{Mock2018} and NdGaO$_3$.\cite{Mock_2019}

Both infrared frequency dependent dielectric function tensor, $\varepsilon(\omega)$, and dielectric loss function tensor, $\varepsilon^{-1}(\omega)$ contain all information on infrared active TO and LO modes, including their directional (anisotropic) properties. Dielectric resonance for electric fields along $\mathbf{\hat{e}}_{l}$ with eigendielectric displacement unit vectors $\mathbf{\hat{e}}_{l} = \mathbf{\hat{e}}_{\mathrm{TO},l}$ define TO mode frequencies. Dielectric loss resonance for electric fields along $\mathbf{\hat{e}}_{l}$ with eigendielectric displacement unit vectors $\mathbf{\hat{e}}_{l} = \mathbf{\hat{e}}_{\mathrm{LO},l}$ define LO frequencies. Note that in the latter case the electric field of an electromagnetic wave does interact with the medium, despite statements to the contrary often found in textbooks.\cite{Kittel2005,Klingshirnbook} The compelling argument is that at the frequency of a LO mode, the lattice polarization (motion)-- the cause of which can only be the electric field of the electromagnetic wave, is oriented so as to compensate the vacuum polarization leading to a vanishing displacement. A set of multiple TO and LO modes, where $l$ may denote a running index, can thus be determined from $\varepsilon(\omega)$ and $\varepsilon^{-1}(\omega)$:\cite{Schubert_2016_LST} 
\begin{subequations}\label{eq:TOLOvectors}
\begin{align}
|\det\{ \varepsilon(\omega=\omega_{\stext{TO},l})\}| &\rightarrow \infty, \\
|\det\{ \varepsilon^{-1}(\omega=\omega_{\stext{LO},l})\}| &\rightarrow \infty, \\
\varepsilon^{-1}(\omega=\omega_{\stext{TO},l})\mathbf{\hat{e}}_{\stext{TO},l} &=0,\\ 
\varepsilon(\omega=\omega_{\stext{LO},l})\mathbf{\hat{e}}_{\stext{LO},l} &=0,
\end{align}
\end{subequations}
 \noindent where det is the determinant. We note without further proof that the total number of TO modes must always equal the total number of LO modes. In the Born and Huang approach,\cite{Born54} the lattice dynamic properties in crystals with arbitrary symmetries are categorized under different electric field $\mathbf{E}$ and dielectric displacement $\mathbf{D}$ conditions.\cite{VFSbook} $\mathbf{E}=0$ and $\mathbf{D}=0$  defines the TO  modes, $\omega_{\mathrm{TO},l}$, associated with dipole moment. $\mathbf{E} \ne 0$ but $\mathbf{D}=0$ defines the LO modes, $\omega_{\mathrm{LO},l}$, identical with the definitions through the dielectric tensor above. $\mathbf{E}\ne 0$ and $\mathbf{D}\ne 0$ defines the so-called limiting frequencies $\omega(\mathbf{\alpha})_{l}$. Here, the limiting frequencies, $\omega(\alpha)_{l}$, depend on the direction of a unit vector within the $\mathbf{a-c}$ plane, $\mathbf{\hat{\alpha}}=\cos{\alpha}\mathbf{\hat{x}}+\sin{\alpha}\mathbf{\hat{y}}$.\cite{SchubertPRB2019} Frequencies $\omega(\mathbf{\alpha})_{l}$ are then obtained from the subtensor within the $\mathbf{a-c}$ plane as follows (T denotes the transpose):
 \begin{equation}
 0=\mathbf{\hat{\alpha}}
\begin{bmatrix}
 \varepsilon_{\mathrm{xx}} & \varepsilon_{\mathrm{xy}}\\
 \varepsilon_{\mathrm{xy}} & \varepsilon_{\mathrm{yy}}
\end{bmatrix}\mathbf{\hat{\alpha}}^{\mathrm{T}}.
 \end{equation}
A physical model must be selected to calculate the effect of the lattice excitation onto the optical properties. In the EDVS approach, $\varepsilon$ is obtained from a sum over all TO mode contributions, added to a high-frequency wavelength independent tensor $\varepsilon_{\infty}$. 
 \begin{equation}\label{eq:epssum}
\varepsilon=\varepsilon_\infty+\sum^{N}_{l=1}\varrho_{\mathrm{TO},l}(\mathbf{\hat{e}}_{\mathrm{TO},l}\otimes\mathbf{\hat{e}}_{\mathrm{TO},l}),
\end{equation}
where $\otimes$ is the dyadic product and $\varrho_{\mathrm{TO},l}$ are wavelength dependent functions. In the EDVLS approach, $\varepsilon^{-1}$ is obtained from a sum over all LO mode contributions, added to a high-frequency wavelength independent inverse tensor $\varepsilon^{-1}_{\infty}$.
\begin{equation}\label{eq:epsinversesum}
\varepsilon^{-1}=\varepsilon^{-1}_\infty-\sum^{N}_{l=1}\varrho_{\mathrm{LO},l}(\mathbf{\hat{e}}_{\mathrm{LO},l}\otimes\mathbf{\hat{e}}_{\mathrm{LO},l}),
\end{equation}
where $\varrho_{\mathrm{TO},l}$ are wavelength dependent functions. Note the minus sign in front of the summation in Eq.~\ref{eq:epsinversesum}, which was chosen to result in real-valued LO mode amplitude parameters in Eq.~\ref{eq:AHLO}.\footnote{We note a misprint in Eq.~3 and Eqs. 9a-9d in Mock~\textit{et al.}\cite{Mock2018}, where a minus sign needs to appear in front of the sum symbols.} Note that both tensors are symmetric, and six complex-valued frequency dependent functions are required to fully render $\varepsilon$ and its inverse, $\varepsilon^{-1}$.  Anharmonic broadened Lorentzian oscillator functions are used to describe wavelength dependent functions in Eqs.~\ref{eq:epssum} and Eqs.~\ref{eq:epsinversesum}.\cite{Mock2018,Mock_2019} 
\begin{equation}\label{eq:AHLO}
\varrho_{k,l} \left(\omega\right)=\frac{A_{k,l}^2-i\Gamma_{k,l}\omega}{\omega^2_{k,l}-\omega^2-i\omega\gamma_{k,l}}.
\end{equation}
\noindent Here, $A_{k,l}$, $\omega_{k,l}$, $\gamma_{k,l}$, and $\Gamma_{k,l}$ denote amplitude, resonance frequency, harmonic broadening, and anharmonic broadening parameter for TO ($k$=``TO'') or LO ($k$=``LO'') mode $l$, respectively, and $\omega$ is the frequency of the driving electromagnetic field. It can be shown that parameters $\Gamma_{k,l}$ vanish when there is no coupling between lattice modes which leads to anharmonic lattice broadening.

A coordinate-invariant generalized dielectric function can be found, which conveniently connects all TO and LO modes within a crystal regardless of its symmetry
\begin{equation}\label{eq:general-eps}
\mathrm{det}\{\varepsilon(\omega)\}=\mathrm{det}\{\varepsilon_\infty\}\prod_{l=1}^{N}\frac{\omega^2_{\stext{LO},l}-\omega^2}{\omega^2_{\stext{TO},l}-\omega^2}.
\end{equation}
This form was described previously by Schubert\cite{Schubert_2016_LST} for crystals with monoclinic or triclinc symmetry, and has similarities with a factorized form described by Berreman and Unterwald\cite{Berreman68} and by Lowndes\cite{Lowndes70} for crystals with orthorhombic and higher symmetry. The Berreman-Unterwald-Lowndes (BUL) form introduces phonon mode lifetime broadening for each TO and LO mode separately. In the same vein, the Schubert form is interpreted with TO and LO broadening parameters which is then valid regardless of crystal symmetry (Schubert-BUL form)
\begin{equation}\label{eq:general-eps-broaded}
\mathrm{det}\{\varepsilon(\omega)\}=\mathrm{det}\{\varepsilon_\infty\}\prod_{l=1}^{N}\frac{\omega^2_{\stext{LO},l}-\omega^2-i\omega\gamma_{\stext{LO},l}}{\omega^2_{\stext{TO},l}-\omega^2-i\omega\gamma_{\stext{TO},l}}.
\end{equation}
The inverse of the Schubert-BUL form is obtained immediately
\begin{equation}\label{eq:generalinv-eps-broaded}
\mathrm{det}\{\varepsilon^{-1}(\omega)\}=\mathrm{det}\{\varepsilon^{-1}_\infty\}\prod_{l=1}^{N}\frac{\omega^2_{\stext{TO},l}-\omega^2-i\omega\gamma_{\stext{TO},l}}{\omega^2_{\stext{LO},l}-\omega^2-i\omega\gamma_{\stext{LO},l}}.
\end{equation}
The usefulness of the inclusion of these forms during the analysis of phonon modes from ellipsometry data has been recently demonstrated for monoclinic $\beta$-Ga$_2$O$_3$,\cite{Schubert2016} CdWO$_4$,\cite{Mock_2017} YSO,\cite{Mock2018} and orthorhombic NdGaO$_3$.\cite{Mock_2019} With $\omega \rightarrow 0$, Eq.~\ref{eq:general-eps} provides the Schubert-Lyddane-Sachs-Teller (S-LST) relationship for crystals with arbitrary symmetry, which relates the DC and high-frequency tensors with all TO and LO modes within a given crystal\cite{Lyddane41,Schubert_2016_LST}
\begin{equation}\label{eq:SLST}
\frac{\mathrm{det}\{\varepsilon(\omega=0)\}}{\mathrm{det}\{\varepsilon_\infty\}}=\prod_{l=1}^{N}\frac{\omega^2_{\stext{LO},l}}{\omega^2_{\stext{TO},l}}.
\end{equation}

In materials with orthorhombic and higher symmetry, frequencies $\omega_{\mathrm{TO,LO},l}$ appear in associated pairs such that a TO mode is always followed by an associated LO mode, where the next mode in order must be a TO mode. This is also known as TO-LO rule.\cite{SchubertIRSEBook_2004} It was found that this order is violated for monoclinic and triclinic symmetries. The resulting phonon order and its relationship with the reststrahlen band and the directional modes was exemplified recently for monoclinic $\beta$-Ga$_2$O$_3$.\cite{SchubertPRB2019} 

Coordinate systems ($\hat{x}$, $\hat{y}$, $\hat{z}$), ($x$, $y$, $z$), and ($a$, $b$, $c$) are needed to describe the optical properties of the monoclinic crystals within the laboratory system of the ellipsometry instrumentation. Laboratory coordinate axes $\hat{x}$, $\hat{y}$, and $\hat{z}$ are associated with the ellipsometer system where $\hat{x}$ is parallel to the plane of incidence and parallel to the sample surface,  $\hat{y}$ is parallel to the sample surface, and $\hat{z}$ is perpendicular to the surface pointing into the sample. The incident wave vector has a positive component in $\hat{x}$ direction. Crystallographic axes of the monoclinic system, ($a$, $b$, $c$), are oriented within the Cartesian sample coordinate system, ($x$, $y$, $z$), such that $z$ is parallel to axis $b$, and axes $a$ and $c$ are within the ($x-y$) plane. Relationships between ($x$, $y$, $z$), and ($a$, $b$, $c$) are shown in Fig.~\ref{fig:unitcell}. Rotation transformations then connect the physical orientation of a given sample mounted onto the sample stage of the ellipsometer with the intrinsic orientations of a given phonon mode within the crystal lattice. Due to factorization according to symmetry and number of elements in the unit cell, 23 TO and 23 LO modes with $\mathrm{A_u}$ symmetry are polarized along vector \textbf{b}. 22 TO and 22 LO modes with $\mathrm{B_u}$ symmetry are polarized within the monoclinic \textbf{a-c} plane.

The orientation of a TO eigendielectric displacement vector with $\mathrm{B_u}$ symmetry relative to $x$ within the ($a$-$c$) plane is denoted by $\alpha_{\mathrm{TO},l}$, then explicitly here 
\begin{subequations}\label{eq:epsmonoall}
\begin{align}
\varepsilon_{\mathrm{xx}} &= \varepsilon_{\infty,\mathrm{xx}}+\sum^{22}_{l=1}\varrho^{\mathrm{B_u}}_{\mathrm{TO},l}\cos^2\alpha_{\mathrm{TO},l},\\
\varepsilon_{\mathrm{xy}} &= \varepsilon_{\infty,\mathrm{xy}}+\sum^{22}_{l=1}\varrho^{\mathrm{B_u}}_{\mathrm{TO},l}\sin\alpha_{\mathrm{TO},l}\cos\alpha_{\mathrm{TO},l},\\
\varepsilon_{\mathrm{yy}} &= \varepsilon_{\infty,\mathrm{yy}}+\sum^{22}_{l=1}\varrho^{\mathrm{B_u}}_{\mathrm{TO},l}\sin^2\alpha_{\mathrm{TO},l},\\
\varepsilon_{\mathrm{zz}} &= \varepsilon_{\infty,\mathrm{zz}}+\sum^{23}_{l=1}\varrho^{\mathrm{A_u}}_{\mathrm{TO},l},\\
\varepsilon_{\mathrm{xy}} &= \varepsilon_{\mathrm{yx}},\\
\varepsilon_{\mathrm{xz}} &= \varepsilon_{\mathrm{zx}} = \varepsilon_{\mathrm{zy}}= \varepsilon_{\mathrm{yz}}=0.
\end{align}
\end{subequations}

The orientation of a LO eigendielectric loss displacement vector with $\mathrm{B_u}$ symmetry relative to $x$ within the ($a$-$c$) plane is denoted by $\alpha_{\mathrm{LO},l}$, then explicitly here
\begin{subequations}\label{eq:epsinvmonoall}
\begin{align}
\varepsilon^{-1}_{\mathrm{xx}} &= \varepsilon^{-1}_{\infty,\mathrm{xx}}-\sum^{22}_{l=1}\varrho^{\mathrm{B_u}}_{\mathrm{LO},l}\cos^2\alpha_{\mathrm{LO},l},\\
\varepsilon^{-1}_{\mathrm{xy}} &= \varepsilon^{-1}_{\infty,\mathrm{xy}}-\sum^{22}_{l=1}\varrho^{\mathrm{B_u}}_{\mathrm{LO},l}\sin\alpha_{\mathrm{LO},l}\cos\alpha_{\mathrm{LO},l},\\
\varepsilon^{-1}_{\mathrm{yy}} &= \varepsilon^{-1}_{\infty,\mathrm{yy}}-\sum^{22}_{l=1}\varrho^{\mathrm{B_u}}_{\mathrm{LO},l}\sin^2\alpha_{\mathrm{LO},l},\\
\varepsilon^{-1}_{\mathrm{zz}} &= \varepsilon^{-1}_{\infty,\mathrm{zz}}-\sum^{23}_{l=1}\varrho^{\mathrm{A_u}}_{\mathrm{LO},l},\\
\varepsilon^{-1}_{\mathrm{xy}} &= \varepsilon^{-1}_{\mathrm{yx}},\\
\varepsilon^{-1}_{\mathrm{xz}} &= \varepsilon^{-1}_{\mathrm{zx}} = \varepsilon^{-1}_{\mathrm{zy}}= \varepsilon^{-1}_{\mathrm{yz}}=0.
\end{align}
\end{subequations}

\subsection{Generalized ellipsometry}

Generalized spectroscopic ellipsometry permits measurements of the optical properties of arbitrarily anisotropic materials,\cite{Schubert1996,SchubertJOSAA13_1996,SchubertADP15_2006} including crystalline materials with monoclinic\cite{JellisonPRB2011CdWO4,Schubert_2016_LST,Schubert2016,Mock_2017,Mock_2017Ga2O3,Sturm_2015,Sturm_2016,Sturm_2017,Mock2018} and triclinic symmetry.\cite{DresselOE2008pentacene} The Mueller matrix is measured, and then compared with model calculations. The Mueller matrix relates the Stokes vector components before and after interaction with a sample, 
\begin{equation}
\left( {{\begin{array}{*{20}c}
 {S_{0} } \hfill \\ {S_{1} } \hfill \\  {S_{2} } \hfill \\  {S_{3} } \hfill \\
\end{array} }} \right)_{\mathrm{output}} =
\left( {{\begin{array}{*{20}c}
 {M_{11} } \hfill & {M_{12} } \hfill \ {M_{13} } \hfill & {M_{14} } \hfill \\
 {M_{21} } \hfill & {M_{22} } \hfill \ {M_{23} } \hfill & {M_{24} } \hfill \\
 {M_{31} } \hfill & {M_{32} } \hfill \ {M_{33} } \hfill & {M_{34} } \hfill \\
 {M_{41} } \hfill & {M_{42} } \hfill \ {M_{43} } \hfill & {M_{44} } \hfill \\
\end{array} }} \right)
\left( {{\begin{array}{*{20}c}
 {S_{0} } \hfill \\ {S_{1} } \hfill \\  {S_{2} } \hfill \\  {S_{3} } \hfill \\
\end{array} }} \right)_{\mathrm{input}}.
\end{equation}
with Stokes vector components defined here by $S_{0}=I_{p}+I_{s}$, $S_{1}=I_{p} - I_{s}$, $S_{2}=I_{45}-I_{ -45}$, $S_{3}=I_{\sigma + }-I_{\sigma - }$. Here, $I_{p}$, $I_{s}$, $ I_{45}$, $I_{-45}$, $I_{\sigma + }$, and $I_{\sigma - }$denote the intensities for the $p$-, $s$-, +45$^{\circ}$, -45$^{\circ}$, right handed, and left handed circularly polarized light components, respectively~\cite{Fujiwara_2007}. As discussed in detail previously,\cite{Schubert1996, Schubert03a,SchubertIRSEBook_2004,Schubert04,Fujiwara_2007,Schubert2016,Mock_2017,Mock2018,Mock_2019} ellipsometry data are compared with model calculated data by best-match regression algorithms. A half-infinite, two phase model with one media being ambient air and the other monoclinic Lu$_{2}$SiO$_{5}$ separated by the planar crystal surface is applied. The angular orientations of the samples relative to their crystallographic orientations are determined together with the wavelength dependencies of the dielectric function tensor elements. Samples with different surface orientations are investigated. Data at multiple sample azimuth orientations and multiple angle of incidences are measured. All sample cuts, azimuthal rotations, and angles of incidence data are best-matched simultaneously for all wavelengths (polyfit), and complex valued functions, $\varepsilon_{\mathrm{xx}}$, $\varepsilon_{\mathrm{yy}}$, $\varepsilon_{\mathrm{xy}}$, and $\varepsilon_{\mathrm{zz}}$ are obtained (wavelength-by-wavelength analysis). All functions are then best-match analyzed using model functions above and varying model parameters. Note that all spectra for $\varepsilon$, $\varepsilon^{-1}$, $\mathrm{det}\{\varepsilon$\}, and $\mathrm{det}\{\varepsilon^{-1}\}$ are evaluated simultaneously to achieve best-match and to find the best-match calculated model parameters.\cite{Mock_2017,Mock2018,Mock_2019}

\section{Experiment}
Two single crystal samples of cerium doped Lu$_2$SiO$_5$ were purchased from MTI Corporation for this investigation. The nominal doping concentration is 0.175~mol$\%$. According to Ning~\textit{et al.}, doping with cerium causes very little change in the lattice structure of the crystal.\cite{Ning2012}  The sample dimensions were each 10~mm~$\times$~10~mm~$\times$~0.5~mm. Crystal orientations of our samples were (001) and (110), respectively, following the axis definition shown in Fig.~\ref{fig:unitcell}. All model calculations were done using WVASE32$^{\mathrm{TM}}$ (J.~A.~Woollam Co.,~Inc.). GSE measurements were performed with two instruments. Data within the infrared spectral range covering approximately 230~cm$^{-1}$ to 1200~$^{-1}$ was acquired with a commercial IR variable angle of incidence spectroscopic ellipsometry (VASE) instrument (J. A. Woollam Co., Inc.). Data within the far-infrared (FIR) spectral range covering approximately 40~cm$^{-1}$ to 500~$^{-1}$ was acquired with an in-house built FIR-VASE instrument.\cite{Kuehne2014} Data was acquired in the Mueller matrix formalism. Mueller matrix data was taken for each sample at two angles of incidence, $\Phi_a=50^{\circ}$, and $70^{\circ}$. Five azimuthal sample orientations for each sample, with the sample rotated clockwise around its normal in 45$^{\circ}$ increments, were measured. Only five azimuthal angles were needed as measurements 180$^{\circ}$ apart are identical and no non-reciprocity effects were observed. All five azimuthal rotations are included in the analysis, but only three rotations for each surface cut are shown in our figures for brevity. Due to our current instrumentation in the FIR spectral region, fourth row elements of the Mueller matrix are only available from the infrared instrument, approximately 230~cm$^{-1}$ and above.

\section{Results and Discussion}

\subsection{DFT Phonon Calculations}\label{dft_phonons}

The phonon frequencies, Born effective charges, and transition dipole components were computed at the $\Gamma$-point of the Brillouin zone using density functional perturbation theory,~\cite{BaroniRMP2001DFTPhonons} as implemented in the Quantum ESPRESSO package. The parameters of the TO modes were obtained from the dynamical matrix computed at the $\Gamma$-point. The parameters of the LO modes were obtained by setting a small displacement from the $\Gamma$-point in order to include the long-range Coulomb interactions of Born effective charges in the dynamical matrix. For $\mathrm{A_u}$ symmetry modes this displacement was in the direction of the crystal vector $\mathbf{b}$. For the $\mathrm{B_u}$ modes, the entire $a$-$c$ plane was probed with a step of 0.1$^{\circ}$, in order to create plots of directional limiting frequencies. The extrema of the dispersion curves for each phonon mode were identified as LO modes if they did not coincide (in terms of phonon frequency and direction) with previously identified TO modes.

The results of the phonon mode calculations for all infrared active modes with $\mathrm{B_u}$ and $\mathrm{A_u}$ symmetry ($\omega_{\mathrm{TO},l}$, $A_{\mathrm{TO},l}$, $\alpha_{\mathrm{TO},l}$, $\omega_{\mathrm{LO},l}$, $A_{\mathrm{LO},l}$, $\alpha_{\mathrm{LO},l}$) are listed in Tabs.~\ref{tab:BuDFT} and \ref{tab:AuDFT}. Note that for modes with $\mathrm{A_u}$ symmetry, all eigenvectors are oriented along direction $\mathbf{b}$ and thus $\alpha_{\mathrm{TO,LO},l}$ are not needed. Values for $\alpha_{\mathrm{TO,LO},}$ for modes with $\mathrm{B_u}$ symmetry are counted relative to the direction of the highest-frequency (TO, LO) mode, and the highest-frequency (TO, LO) mode is counted relative to axis $\mathbf{a}$ within the $\mathbf{a-c}$ plane. Renderings of atomic displacements for each mode were prepared using XCrysDen~\cite{XCrysDen} running under Silicon Graphics Irix 6.5, and are shown in Figs.~\ref{fig:TOphonons} and~\ref{fig:LOphonons}.

\begin{table*}
\caption{\label{tab:BuDFT} Phonon mode parameters for B$_{\mathrm{u}}$ symmetry modes obtained by DFT. Units are reciprocal centimeters (cm$^{-1})$, Debye (D), Angstrom (\AA), angular degrees ($^{\circ}$), and atomic mass units (amu). Parameters for the angular orientation are relative to Mode 1, defined from the unit cell direction $a$ as 28.76$^{\circ}$ and 32.70$^{\circ}$ for the TO and LO modes, respectively.}
\begin{ruledtabular}
\begin{tabular}{{l}{c}{c}{c}{c}{c}{c}}
Mode & $\omega_{\mathrm{TO},l}$ (cm$^{-1}$) & $A^2_{\mathrm{TO},l}$ [(D/\AA)$^2$/amu] & $\alpha_{\mathrm{TO},l}$ ($^\circ$) & $\omega_{\mathrm{LO},l}$ (cm$^{-1}$) & $A^2_{\mathrm{LO},l}$ [(D/\AA)$^2$/amu] & $\alpha_{\mathrm{LO},l}$ ($^\circ$)\\
\hline
1  & 928.14 & 78.497 & 0.00   & 1009.14  & 124.524 & 0.00   \\
2  & 885.18 & 46.136 & 101.26 & 953.64 & 111.495 & 91.70  \\
3  & 857.46 & 23.077 & 86.98  & 866.22  & 3.329   & 77.03  \\
4  & 840.80 & 8.042  & 93.72  & 843.15  & 0.782   & 95.19  \\
5  & 559.52 & 40.240 & 24.87  & 636.60  & 59.184  & 31.59  \\
6  & 526.89 & 18.174 & 50.00  & 529.39  & 9.603   & 146.65 \\
7  & 505.95 & 6.360  & 114.06 & 507.61  & 0.769   & 49.67  \\
8  & 500.95 & 21.230 & 177.09 & 547.00  & 23.385  & 113.66 \\
9  & 460.29 & 8.576  & 75.90  & 472.56  & 6.020   & 97.72  \\
10 & 392.41 & 4.013  & 72.33  & 410.18  & 14.704  & 116.56 \\
11 & 360.20 & 11.746 & 130.00 & 388.98  & 5.542   & 149.85 \\
12 & 296.51 & 23.377 & 146.41 & 324.30  & 9.928   & 37.47  \\
13 & 275.34 & 22.485 & 68.55  & 333.68  & 5.730   & 109.76 \\
14 & 270.29 & 2.137  & 73.00  & 270.74  & 0.011   & 84.58  \\
15 & 233.68 & 19.202 & 58.12  & 249.85  & 1.026   & 45.97  \\
16 & 204.62 & 8.804  & 3.59   & 231.77  & 3.336   & 150.36 \\
17 & 183.91 & 20.808 & 139.75 & 202.51  & 0.475   & 109.24 \\
18 & 161.70 & 1.971  & 137.48 & 165.57  & 0.338   & 92.76  \\
19 & 149.48 & 13.658 & 98.44  & 160.36  & 0.214   & 71.01  \\
20 & 104.13 & 0.573  & 59.41  & 105.24  & 0.042   & 52.58  \\
21 & 76.58  & 1.147  & 106.22 & 79.07   & 0.057   & 110.39 \\
22 & 64.04  & 0.236  & 143.65 & 64.70   & 0.014   & 163.56\\
\end{tabular}
\end{ruledtabular}
\end{table*}

\begin{table*}
\caption{\label{tab:AuDFT} Same as Tab.~\ref{tab:BuDFT} for modes with A$_{\mathrm{u}}$ symmetry.}
\begin{ruledtabular}
\begin{tabular}{{l}{c}{c}{c}{c}}
Mode & $\omega_{\mathrm{TO},l}$ (cm$^{-1}$)& $A^2_{\mathrm{TO},l}$ [(D/\AA)$^2$/amu] & $\omega_{\mathrm{LO},l}$ (cm$^{-1}$) & $A^2_{\mathrm{LO},l}$ [(D/\AA)$^2$/amu] \\
\hline
1  & 920.59 & 1.3255  & 963.53 & 113.2478 \\
2  & 894.12 & 33.3283 & 919.52 & 1.1628   \\
3  & 868.09 & 36.8330 & 880.20 & 3.0297   \\
4  & 848.54 & 5.6808  & 850.13 & 0.5144   \\
5  & 596.40 & 13.2935 & 620.52 & 23.081   \\
6  & 552.78 & 5.6431  & 560.76 & 4.9752   \\
7  & 523.74 & 4.0507  & 529.53 & 3.3011   \\
8  & 504.19 & 3.2719  & 508.11 & 1.8249   \\
9  & 427.37 & 5.5023  & 444.35 & 11.8087  \\
10 & 409.18 & 2.3907  & 413.43 & 1.7162   \\
11 & 362.48 & 0.0001  & 362.48 & 0.0001   \\
12 & 338.47 & 7.6803  & 378.18 & 12.698   \\
13 & 319.26 & 15.4837 & 331.82 & 0.6623   \\
14 & 280.78 & 7.3042  & 291.80 & 1.5379   \\
15 & 238.45 & 6.3969  & 256.70 & 2.1276   \\
16 & 224.07 & 4.8942  & 231.21 & 0.3554   \\
17 & 202.08 & 22.1504 & 217.30 & 0.352    \\
18 & 185.93 & 2.6995  & 187.23 & 0.0346   \\
19 & 163.53 & 0.4978  & 163.99 & 0.0194   \\
20 & 136.39 & 2.7855  & 140.03 & 0.1483   \\
21 & 106.62 & 0.6106  & 107.71 & 0.0375   \\
22 & 90.79  & 0.4051  & 91.67  & 0.0267   \\
23 & 81.52  & 0.4565  & 82.49  & 0.0228  \\
\end{tabular}
\end{ruledtabular}
\end{table*}

\subsection{Mueller matrix analysis}

\begin{figure*}[hbt]
\centering
\includegraphics[width=\linewidth]{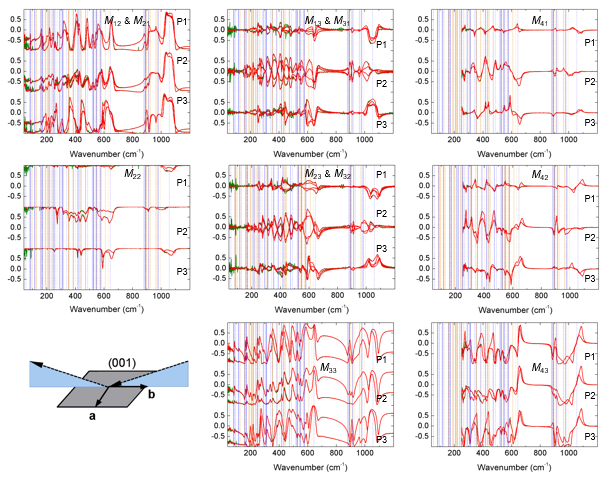}
\caption{\label{fig:MM001} Lu$_2$SiO$_5$ GSE data at $\Phi_a=50^{\circ}$ and $70^{\circ}$ angle of incidence:    Dotted green lines (experiment); Solid red lines (best match model calculated). Data are presented in the Mueller matrix formalism. All data are normalized to element M$_{11}$. The sample surface is (001), with best match calculated Euler angle parameters $\theta=80.8(4)$ and $\psi=-11.1(5)$. Data are shown for three azimuths: P1 [$\varphi=-84.4(2)^{\circ}$]; P2 [$\varphi=-39.4(2)^{\circ}$]; P3 [$\varphi=5.5(8)^{\circ}$]. TO and LO modes are indicated by solid and dotted lines, respectively, for $\mathrm{B_u}$ symmetry (blue) and $\mathrm{A_u}$ symmetry (brown).}
\end{figure*}

\begin{figure*}[hbt]
\centering
\includegraphics[width=\linewidth]{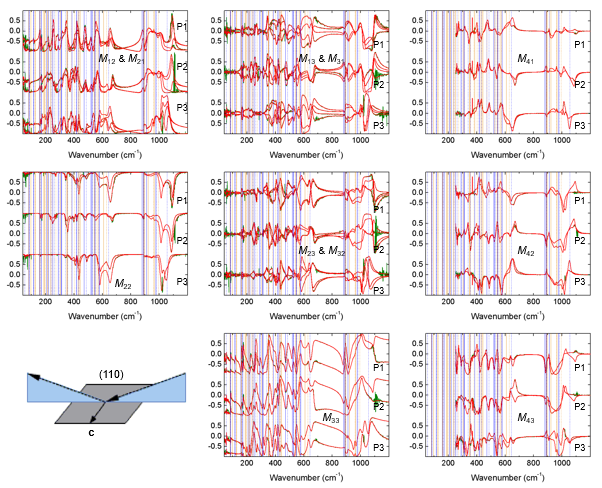}
\caption{\label{fig:MM110}Same as Fig.~\ref{fig:MM001} for (110) Lu$_2$SiO$_5$ with $\theta=42.9(3)$ and $\psi=15.1(8)$. P1 [$\varphi=-92.0(7)^{\circ}$]; P2 [$\varphi=-47.7(4)^{\circ}$]; P3 [$\varphi=-2.7(4)^{\circ}$].}
\end{figure*}

Experimental Mueller matrix data and the best-match model data are shown in Figs. \ref{fig:MM001} and \ref{fig:MM110} for both samples studied. Each unique Mueller matrix element is shown in its own pane and arranged by the corresponding matrix indices. Each pane shows three azimuthal positions denoted by P1, P2, and P3 with two angles of incidence each (50$^\circ$ and 70$^\circ$). Data sets symmetric in indices are plotted together and as denoted in the corresponding panes. All Mueller matrix elements are normalized to $M_{11}$. Element $M_{44}$ cannot be measured with our current instrumentation and is therefore not presented. Data obtained within the FIR range (40~cm$^{-1}$ to 500~cm$^{-1}$) and data within the IR range (500~cm$^{-1}$ to 1200~cm$^{-1}$) are shown for all elements excluding the fourth row elements. Due to limitations of our current FIR instrumentation, only IR data (250~cm$^{-1}$ to 1200~cm$^{-1}$) is shown for all fourth row elements. 

In-plane anisotropy is seen in Figs.~\ref{fig:MM001} and~\ref{fig:MM110} in the off-block diagonal elements ($M_{13}$, $M_{23}$, $M_{14}$, and $M_{24}$). For isotropic samples, these elements are zero valued across the entire spectra. Each of these elements are strongly impacted by the azimuthal rotations. To show correlation between all Mueller matrix elements and the extracted dielectric function tensor elements, the frequencies of all TO and LO phonon modes with $\mathrm{A_u}$ and $\mathrm{B_u}$ symmetries are shown as horizontal lines in Figs.~\ref{fig:MM001} and \ref{fig:MM110}. Our polyfit data for each wavelength included up to 792 independent data points from the multiple cuts, azimuthal rotations, and angles of incidence. In this wavelength-by-wavelength analysis, 14 independent parameters are varied. Of these 14, eight are the real and imaginary parts of the dielectric function tensor elements ($\varepsilon_{\mathrm{xx}}$, $\varepsilon_{\mathrm{yy}}$, $\varepsilon_{\mathrm{xy}}$, and $\varepsilon_{\mathrm{zz}}$). The remaining six variables are two sets of three Euler angles used to describe the sample surface and orientation, independent of wavelength. The resultant Mueller matrices of this polyfit are shown in Figs. \ref{fig:MM001} and \ref{fig:MM110} as the solid red lines. The dielectric tensor elements found in this fit are shown in Fig. \ref{fig:tensor} as the dotted green lines. Overall, there is excellent agreement between the experimental and calculated Mueller matrix data. Longer wavelength data does become noisier as a result of lower source intensity. While these LSO samples were doped with cerium, no free charge carrier effects are detected in our data. 

\subsection{Dielectric tensor analysis}

\begin{figure*}[hbt]
\centering
\includegraphics[width=\textwidth]{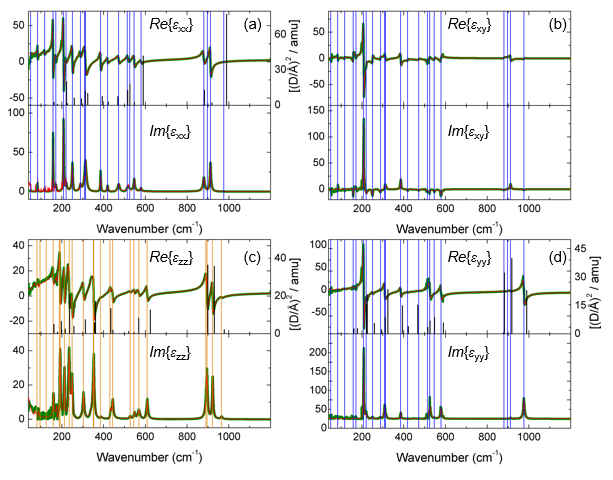}
\caption{\label{fig:tensor}Lu$_2$SiO$_5$ dielectric function tensor spectra: Green dotted lines (GSE);  Red solid lines (best match EDVLS model); Vertical blue lines (B$_\mathrm{u}$ symmetry TO modes); Vertical orange lines A$_\mathrm{u}$ symmetry TO modes); Vertical black bars (DFT transition dipole moments).}
\end{figure*}

\begin{figure*}[hbt]
\centering
\includegraphics[width=.9\textwidth]{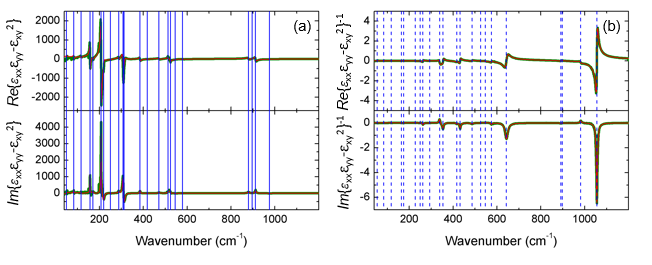}
\caption{\label{fig:det}(a) Coordinate invariant generalized dielectric function: Green dotted lines (GSE);  Red solid lines (best match Schubert-BUL form); Vertical blue lines (all TO modes). (b) Coordinate invariant inverse generalized dielectric function: Green dotted lines (GSE);  Red solid lines (best match inverse Schubert-BUL form); Vertical dashed blue lines (all LO modes).}
\end{figure*}

\begin{figure*}[hbt]
\centering
\includegraphics[width=.9\linewidth]{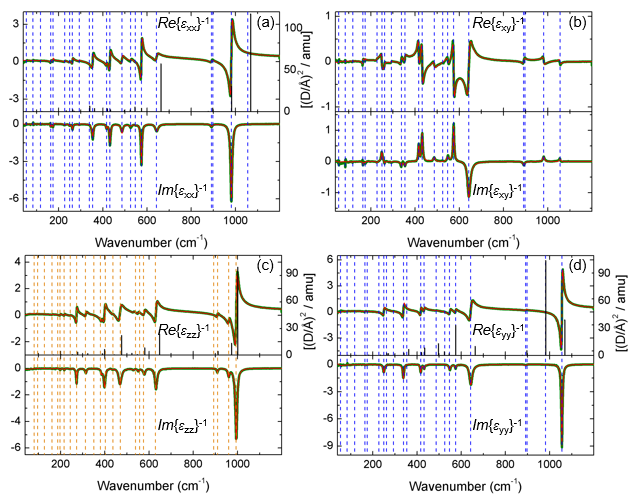}
\caption{\label{fig:inversetensor}Same as for Fig.~\ref{fig:tensor} for the inverse dielectric tensor: Vertical dashed blue lines ($\mathrm{B_u}$ symmetry LO modes); Vertical dashed orange lines ($\mathrm{A_u}$ symmetry TO modes).}
\end{figure*}
The real and imaginary parts of the dielectric function tensor elements found during the polyfit ($\varepsilon_{\mathrm{xx}}$, $\varepsilon_{\mathrm{xy}}$, $\varepsilon_{\mathrm{yy}}$, and $\varepsilon_{\mathrm{zz}}$) are shown in Fig.~\ref{fig:tensor} as green dotted lines. These tensor elements were then translated into the inverse dielectric function tensor elements ($\varepsilon^{-1}_{\mathrm{xx}}$, $\varepsilon^{-1}_{\mathrm{xy}}$, $\varepsilon^{-1}_{\mathrm{yy}}$, and $\varepsilon^{-1}_{\mathrm{zz}}$) as shown in Fig.~\ref{fig:inversetensor} as the green dotted lines again. Similarly, the determinant and inverse determinant elements ($\varepsilon_{\mathrm{xx}}\varepsilon_{\mathrm{yy}}-\varepsilon_{\mathrm{xy}}^2$) and inverse determinant (($\varepsilon_{\mathrm{xx}}\varepsilon_{\mathrm{yy}}-\varepsilon_{\mathrm{xy}}^2)^{-1}$) shown in Fig.~\ref{fig:det} are derived from the polyfit elements.

From these dielectric tensor elements, phonon modes can be observed. TO mode resonant frequencies occur at the maxima of the imaginary parts of the dielectric function tensor elements and the determinant \cite{Schubert_2016_LST}. Similarly, LO mode resonant frequencies occur at the maxima of the imaginary parts of the inverse dielectric function elements and inverse determinant.\cite{Mock_2017} In Figs.~\ref{fig:tensor} and \ref{fig:inversetensor}, panels (a), (b), and (d) have common peaks where we identify 22 TO and LO mode pairs with $\mathrm{B_u}$ symmetry (in the $a$- plane). Likewise, the peaks of panel (c) show 23 TO and LO mode pairs along the $\mathrm{A_u}$ symmetry axis.

\subsection{Phonon mode analysis}

\subsubsection{Modes with $\mathrm{B_u}$ symmetry in the $a$-$c$ plane}

\paragraph{TO mode parameter determination} By using a set of anharmonically broadened Lorentzian oscillators we derive the best-match model calculations shown in Figs. \ref{fig:tensor} and \ref{fig:inversetensor} as the solid red lines. The best-match TO model parameters are detailed in Tab. \ref{Table:ExpBu}. Parameters included in this table are amplitude ($A_{\mathrm{TO},l}$), frequency ($\omega_{\mathrm{TO},l}$), broadening ($\gamma_{\mathrm{TO},l}$), anharmonic broadening ($\Gamma_{\mathrm{TO},l}$), and eigenvector direction ($\alpha_{\mathrm{TO},l}$) for all TO modes ($l=1...22$) with $\mathrm{B_u}$ symmetry.
\begin{table*}\centering 
\caption{\label{Table:ExpBu}GSE parameters for $\mathrm{B_u}$ symmetry TO and LO modes. $\alpha_{\mathrm{TO}}$ and $\alpha_{\mathrm{LO}}$ are relative to Mode 1, defined from direction \textbf{$x$} as 91.63~$^{\circ}$ and -89.34~$^{\circ}$ for the TO and LO mode, respectively.}
\begin{ruledtabular}
\begin{tabular}{{l}|{c}{c}{c}{c}{c}{c}{c}{c}{c}{c}}
Mode & \scriptsize{$\omega_{\mathrm{TO}}$(cm$^{-1}$)}&\scriptsize{$\omega_{\mathrm{LO}}$(cm$^{-1}$)}& \scriptsize{$\gamma_{\mathrm{TO}}$(cm$^{-1}$)}&\scriptsize{$\gamma_{\mathrm{LO}}$(cm$^{-1}$)}&\scriptsize{$A_{\mathrm{TO}}$(cm$^{-1}$)}&\scriptsize{$\Gamma_{\mathrm{TO}}$(cm$^{-1}$)}&\scriptsize{$\alpha_{\mathrm{TO}}$($^\circ$)}&\scriptsize{$A_{\mathrm{LO}}$(cm$^{-1}$)}&\scriptsize{$\Gamma_{\mathrm{LO}}$(cm$^{-1}$)}&\scriptsize{$\alpha_{\mathrm{LO}}$($^\circ$)}\\
\hline
1 & 975.(6) & 1055.5(0) & 8.(1) & 6.5(0) & 6(5)0 & -5.(6) & (0) & 24(8) & 0.(1) & 0 \\
2 & 911.(4) & 981.3(7) & 6.(9) & 9.(3) & 5(0)0 & -7.(5) & 10(3) & 23(6) & -0.(3) & 87.(9) \\
3 & 89(6) & 89(7) & 2(3) & 1(8) & 7(9) & 1(0) & 7(5) & (1) & 0.(8) & 3.(8) \\
4 & 879.(9) & 890.(8) & 9.(7) & 9.(7) & 3(9)0 & 1(7) & 8(3) & 4(5) & -0.(1) & 11(5)  \\
5 & 578.(9) & 643.0(4) & 7.(2) & 15.1(2) & 3(7)0 & 1(0) & 1(8) & 165.(5) & -4.(3) & 152.(0) \\
6 & 545.(9) & 574.7(2) & 7.(2) & 6.(8) & 2(5)0 & -6.(8) & 7(5) & 11(8) & 0.(5) & 69.(9)\\
7 & 524.(7) & 546.(9) & (8) & 1(0) & 3(8)0 & -(2)9 & 2.(5) & 5(9) & -0.(9) & 1(3) \\
8 & 51(3) & 52(5) & 1(8) & 25$^\textrm{a}$ & 2(6)0 & -(7)0 &5(9) & 4(2) & 1.(3) & 9(8)\\
9 & 471.(9) & 487.(0) & 12.(7) & 12.(7) & 2(5)0 & 0.(8) & 7(6) & 5(6) & 0.(5) & 7(9) \\
10 & 418.(5) & 432.0(5) & 5.(6) & 6.5(7) & 1(4)0 & (0) & 6(9) & 7(6) & -1.(0) & 5(9)\\
11 & 385.(1) & 416.(5) & 3.(6) & 5.(6) & 2(6)0 & (2) & 12(9) & 5(4) & -0.(3) & 2(9)\\
12 & 312.(4) & 354.(4) & 11.(4) & 8.4(9) & 3(9)0 & -(1) & 10(0) & 5(7) & -0.(2) & 9(7)\\
13 & 308.(5) & 339.0(3) & 6.(9) & 7.3(3) & 2(7)0 & -(2)4 & 1.(8) & 6(2) & 2.(1) & 17(4)\\
14 & 28(9) & 293.(7) & (9) & (6) & 1(7)0 & 1(4) & 5(5) & 1(2) & 0.(4) & 10(1)\\
15 & 250.(2) & 250.(1) & 5.(2) & 7.(3) & 22(9) & -(3)0 & 7(0) & 4(3) & 0.(4) & 1(9)\\
16 & 220.(2) & 227.(8) & (8) & (7) & 2(0)0 & -(2)0 & 4(0) & 1(4) & 0.(0) & 11(4)\\
17 & 207.6(6) & 262.2(8) & 2.9(8) & 4.(9) & 40(7) & -1(0) & 142.(5) & 2(4) & -0.(2) & 9(9) \\
18 & 171.(1) & 175.(1) & (7) & (4) & 1(0)0 & (3)0 & 10(7) & 1(2) & 0.(5) & 10(1)\\
19 & 11(9) & 12(0) & 1(3) & (6) & (7) & 2(6) & 3(6) & 4.(2) & 0.(2) & 11(6)\\
20 & 157.4(4) & 164.(2) & 2.(1) & 2.(9) & 16(0) & -(1)6 & 7(5) & 1(1) & -0.(1) & 11(9)\\
21 & 83.(0) & 83.(8) & 1.(3) & 1.(3) & 4(1) & -1(1) & 1(1)0 & 2.(3) & -0.(1) & 3(1) \\
22 & 5(0)$^\textrm{a}$ & 5(3)$^\textrm{a}$ & 1(0) & 1(9) & 1(0) & (1)0 & 9(2) & 1(3) & -2.(3) & 16(9) \\
		\end{tabular}
\end{ruledtabular}
\begin{flushleft}
\footnotesize{$^\textrm{a}${Mode parameters fit in a local region, held constant in full spectral fit procedure.}}\\
\end{flushleft}
\end{table*}

\paragraph{TO eigendielectric displacement vectors}

\begin{figure*}[hbt]
\centering
\includegraphics[width=\linewidth]{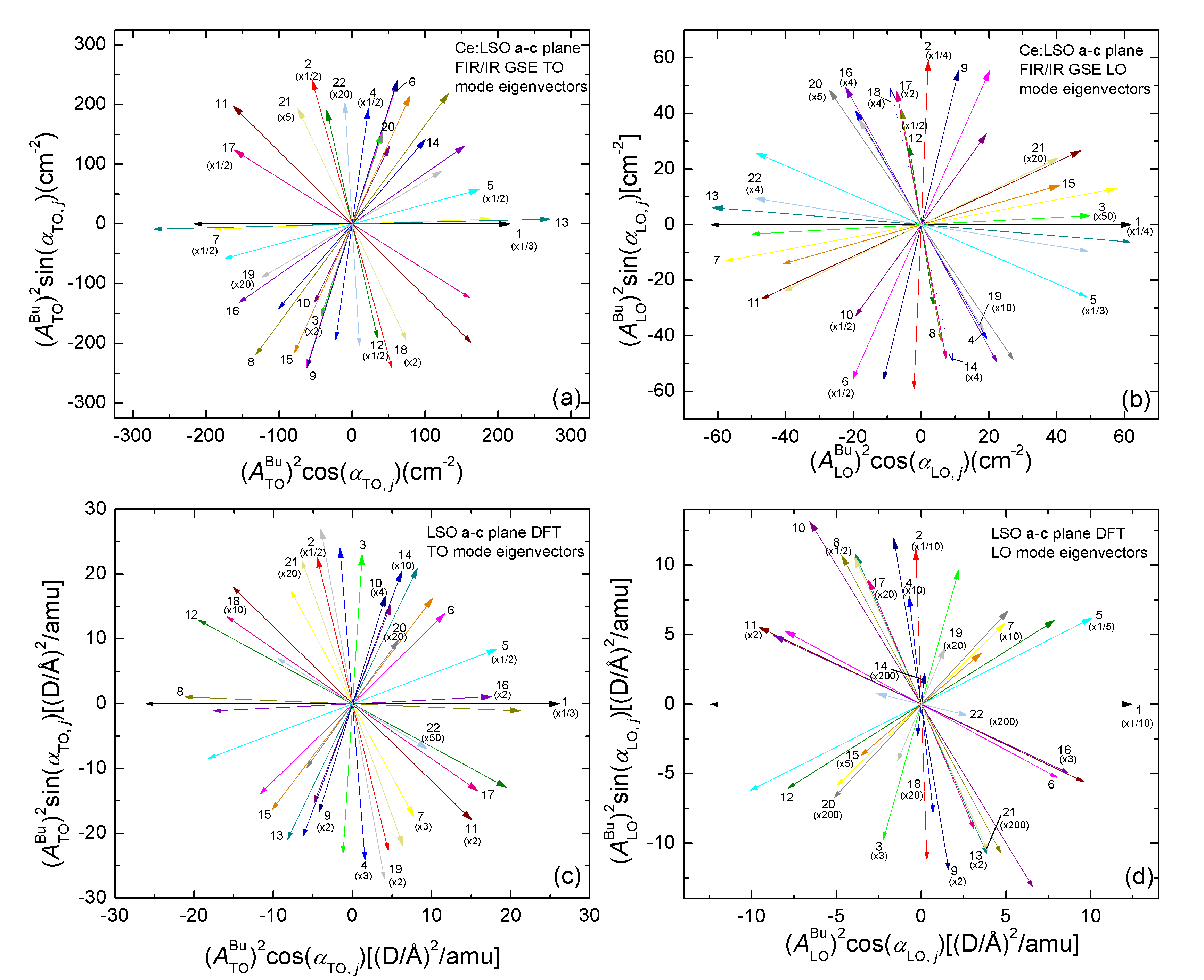}
\caption{\label{fig:TOvectors} (a) Schematic representation of the eigendielectric displacement vectors with GSE analysis determined amplitude $A_{\mathrm{TO},l}^{\mathrm{Bu}}$ and orientation angle $\alpha_{\mathrm{TO},l}$ (with respect to the crystal direction $a$) of TO modes with $\mathrm{B_u}$ symmetry within the $a$-$c$ plane. (c) DFT calculated infrared transition dipoles (intensities) of TO modes with $\mathrm{B_u}$ symmetry.  (b) Schematic representation of the eigendielectric displacement loss vectors with GSE analysis determined amplitude $A_{\mathrm{LO},l}^{\mathrm{Bu}}$ and orientation angle $\alpha_{\mathrm{LO},l}$ (with respect to the crystal direction $a$) of LO modes with $\mathrm{B_u}$ symmetry within the $a$-$c$ plane. (d) DFT calculated infrared transition dipoles (intensities) of LO modes with $\mathrm{B_u}$ symmetry.}
\end{figure*}

Figure~\ref{fig:TOvectors} is a vector representation of the amplitude and polarization direction parameters ($A^{\mathrm{Bu}}_{\mathrm{TO},l}$ and $\alpha_{\mathrm{TO},l}$) within the $a$-$c$ plane. Here, GSE (a), (b) model data are compared to DFT derived data (c), (d). TO vectors are shown in panels (a) and (c) and LO vectors are in (b) and (d). Note that small amplitude modes are enlarged by factors as shown. There is generally good agreement between the GSE and DFT data, particularly for modes with large amplitudes, such as mode 1, 5, 11, or 13, for example. Some modes appear similar but are numbered differently between the two data sets. This may happen as a result of a slightly different frequency being found in GSE than DFT and hence a different number is being assigned to the mode. For instance, this happens in the TO ((a) and (c)) modes 3 and 4. There is noticeable disagreement between other modes. This lack of agreement may be attributed to a weaker mode amplitude parameter or being too close in frequency to another mode feature.

\paragraph{LO mode parameter determination} Blue solid lines in Figs.~\ref{fig:tensor} and ~\ref{fig:inversetensor} indicate the resulting best-match model calculations obtained from Eq.~\ref{eq:epsmonoall} using a second independent set of anharmonically broadened Lorentzian oscillators (LO mode summation). We find excellent agreement between our wavelength-by-wavelength and model calculated $\varepsilon^{-1}$ and $\varepsilon$. All best match LO model parameters are summarized in Tab.~\ref{Table:ExpBu} including amplitude ($A_{\mathrm{LO},l}$), frequency ($\omega_{\mathrm{LO},l}$), broadening ($\gamma_{\mathrm{LO},l}$), anharmonic broadening ($\Gamma_{\mathrm{LO},l}$), and eigenvector direction ($\alpha_{\mathrm{LO},l}$) for all LO modes ($l=1...22$) with $\mathrm{B_u}$ symmetry. Frequencies of the LO modes are indicated by dotted vertical blue lines in Figs.~\ref{fig:MM001},~\ref{fig:MM110},~\ref{fig:det}, and~\ref{fig:inversetensor} which align with the features observed in the data and the extrema seen in the imaginary part of the inverse dielectric tensor. 

\subsubsection{Modes with A$_\mathrm{u}$ symmetry along the crystal direction $b$}
GSE results of mode parameters for A$_\mathrm{u}$ modes are listed in Tab.~\ref{Table:ExpAu}. The dielectric function and inverse dielectric functions are shown in Figs.~\ref{fig:tensor}(c) and~\ref{fig:inversetensor}(c), respectively. Red lines show our best match model calculation using 23 anharmonic Lorentzian oscillators. Parameter sensitivity is critical for modes which occur in close wavelength proximity and/or possess small amplitudes. Weak modes with small amplitude parameters may become subsumed by stronger modes during the regression analysis. Therefore, manual parameter adjustments and limited parameter regions for some parameters were used to reach best-match calculated model parameters. These adjustments are noted in Tab.~\ref{Table:ExpAu} accordingly. Modes 4, 11, 13, 15, and 19~-~23 possess small splitting between their TO and LO frequencies. We have treated those hence as impurity-like vibrational modes as discussed recently for wurtzite-structure GaN.\cite{KasicPRB62_2000} Accordingly, their TO and LO frequencies are listed equal in Tabl.~\ref{Table:ExpAu}. We note that modes 11 and 15 are observed within the TO-LO bands formed by modes 12 and 16, respectively, while all other impurity-like vibrational modes are located outside such bands. Frequencies of TO modes with A$_\mathrm{u}$ symmetry are indicated by vertical solid brown lines in Figs.~\ref{fig:MM001},~\ref{fig:MM110}, and~\ref{fig:tensor} while frequencies of LO modes with A$_\mathrm{u}$ symmetry are indicated by vertical dotted brown lines in Figs.~\ref{fig:MM001},~\ref{fig:MM110}, and ~\ref{fig:inversetensor}. Table~\ref{tab:AuDFT} lists DFT calculated frequencies and amplitude parameters for A$_{\mathrm{u}}$ modes, and overall a good agreement is observed with our GSE results. Modes 19-23 possess small TO-LO mode splittings, with mode 20 as small exception being predicted as a band with slightly larger TO-LO splitting. Likewise, modes 4 and 11 are predicted with very small TO-LO splitting, in agreement with our observation. Note that mode 11 is located within the TO-LO band of mode 12, consistent with our GSE results. Modes 13 and 15 are predicted with somewhat larger TO-LO splitting than observed. We note that mode 15 is also observed to be located within mode 16 in GSE, while in DFT modes 15 and 16 are separated.

\begin{table*}\centering 
\caption{\label{Table:ExpAu} Same as for Tab.~\ref{Table:ExpBu} for $\mathrm{A_u}$ symmetry modes.}
\begin{ruledtabular}
\begin{tabular}{{l}|{c}{c}{c}{c}{c}{c}{c}{c}}
Mode & $\omega_{\mathrm{TO}}$(cm$^{-1}$) & $\omega_{\mathrm{LO}}$(cm$^{-1}$)& $\gamma_{\mathrm{TO}}$(cm$^{-1}$) & $\gamma_{\mathrm{LO}}$(cm$^{-1}$)& $A_{\mathrm{TO}}$(cm$^{-1}$) & $\Gamma_{\mathrm{TO}}$(cm$^{-1}$)& $A_{\mathrm{LO}}$(cm$^{-1}$)& $\Gamma_{\mathrm{LO}}$(cm$^{-1}$) \\
\hline
1  & 963.9(5) & 994.6(3) & 10.1(9) & 9.91(7) & 115.(6) & -2.9(3) & 230.0(6) & 1.2(3)     \\
2  & 922.1(0) & 959.7(3)  & 6.4(9)  & 8.1(5)  & 386.(7) & -9.(9)  & 64.(1)   & -0.8(8)    \\
3  & 895.1(2) & 909.4(6)  & 2.(5)   & 5.8(9)  & 7(9)    & -1(1)   & 40.4(5)  & 0.1(8)    \\
4  & 894.2(3) & 894.2(3)  & 9.(8)   & 2.(5)   & 48(3)   & 3(3)    & 1.(0)    & 0.0(1)    \\
5  & 608.2(3) & 631.3(9)  & 8.4(9)  & 11.8(4) & 238.(8) & -6.(2)  & 108.(0)  & -1.1(9)    \\
6  & 568.6(6) & 579.2(0)  & 13.(1)  & 11.(0)  & 20(0)   & 2.(5)   & 53.(6)   & 0.7(3)     \\
7  & 547.(7)  & 550.(8)   & 7.(5)   & 7.(2)   & 12(0)   & 4.(1)   & 26.(3)   & -0.0(7)    \\
8  & 527.(4)  & 529.(0)   & 8.(1)   & 8.(1)   & 8(1)    & 1.(2)   & 21.(2)   & -0.0(4)    \\
9  & 442.6(3) & 468.7(5)  & 7.(5)   & 15.0(8) & 19(4)   & -1(7)   & 92.(5)   & -1.4(7)    \\
10 & 434.(0)  & 436.(4)   & 6.(6)   & 6.(0)   & 12(4)   & 1.(7)   & 13.(7)   & 0.0(7)     \\
11 & 386.5(8) & 386.5(8)  & 7.(2)   & 5.(5)   & 43.(9)  & -3.(6)  & 24.(9)   & -0.4(3)    \\
12 & 352.9(8) & 399.2(9)  & 7.5(4)  & 9.2(5)  & 28(8)   & 6(9)    & 71.(8)   & 1.4(6)     \\
13 & 346.(0)  & 346.(0)   & 26.(7)  & 23.(9)  & 20(8)   & -9(4)   & (1)      & 0.4(2)     \\
14 & 303.5(3) & 314.6(7)  & 9.3(1)  & 7.2(6)  & 205.(3) & 9.(8)   & 28.(7)   & 0.1(4)     \\
15 & 249.6(2) & 249.6(2)  & 5.8(2)  & 8.(9)   & 177.(3) & 3(0)    & 10.6(6)  & 0.1(2)    \\
16 & 234.0(7) & 272.3(0)  & 9.1(6)  & 4.4(3)  & 288.(6) & -2(9)   & 37.5(1)  & -0.2(6)    \\
17 & 212.2(0) & 217.2(0)  & 5.0(1)  & 3.7(7)  & 177.(6) & 3(1)    & 10.6(4)  & 0.0(1)    \\
18 & 191.6(8) & 198.2(2)  & 5.6(5)  & 5.1(1)  & 211.(3) & 2(6)    & 10.8(8)  & -0.0(2)   \\
19 & 175.7(3) & 175.7(3)  & 3.(5)   & 2.(2)   & 3(6)    & 2(6)    & 2.8(5)   & 0.0(7)    \\
20 & 159.(0) & 159.(0)  & 5.(3)   & 5.(7)   & 10(3)   & 3.(1)   & 6.3(5)   & -0.0(5)   \\
21 & 88.5$^a$  & 88.5$^a$  & 1.(5)  & 1.(4)   & 1(9)    & 2.(8)   & (1)      & -0.0(7) \\
22 & 69$^a$   & 69$^a$    & 5.(9)   & 7.(9)   & 2(5)    & -1(2)   & (1)      & -0.(2) \\
23 & 32$^a$   & 32$^a$    & 2(7)    & 3(0)  & 12(7)   & -6(6)   & 10.(9)   & -0.2(6) 
		\end{tabular}
        \end{ruledtabular}
\begin{flushleft}
\footnotesize{$^a${Mode parameters fit in a local region, held constant in full spectral fit procedure.}}\\
\end{flushleft}
\end{table*}

\subsubsection{TO-LO rule}\label{sec:TOLOrule} As pointed out by Mock~\textit{et al.},\cite{Mock_2019} the Schubert-BUL form (Eq.~\ref{eq:general-eps-broaded}) can be used to identify violations of the TO-LO order. Negative imaginary parts occur in frequency regions of TO-LO bands (inner modes, ``+'') nested within larger TO-LO bands (outer modes, ``-''). For example, a mode sequence of [TO$_-$ ... [TO$_{1,+}$,LO$_{1,+}$]...[TO$_{n,+}$,LO$_{n,+}$]  ... LO$_+$] will show negative imaginary parts in frequency bands [TO$_{1,+}$,LO$_{1,+}$], [TO$_{2,+}$,LO$_{2,+}$], ... and [TO$_{n,+}$,LO$_{n,+}$]. It is noted that this form does not represent a measurable dielectric function and represents an indicator of physical properties rather than representing a physical property itself. Hence, a negative imaginary part is not prohibitive. Here, we observe such occurrences between TO-1 and LO-2, in the very narrow range between TO-7 and LO-8, between TO-12 and LO-13, and between TO-15 and LO-15 and TO-16 and LO-16. These frequency regions are identical with bands of total reflection and the formation of inner and outer modes as discussed below. Note that the TO-LO rule holds true for all A$_\mathrm{u}$ modes.

\subsubsection{Phonon mode order and directional modes}
\begin{figure*}[hbt]
\centering
\includegraphics[width=.9\linewidth]{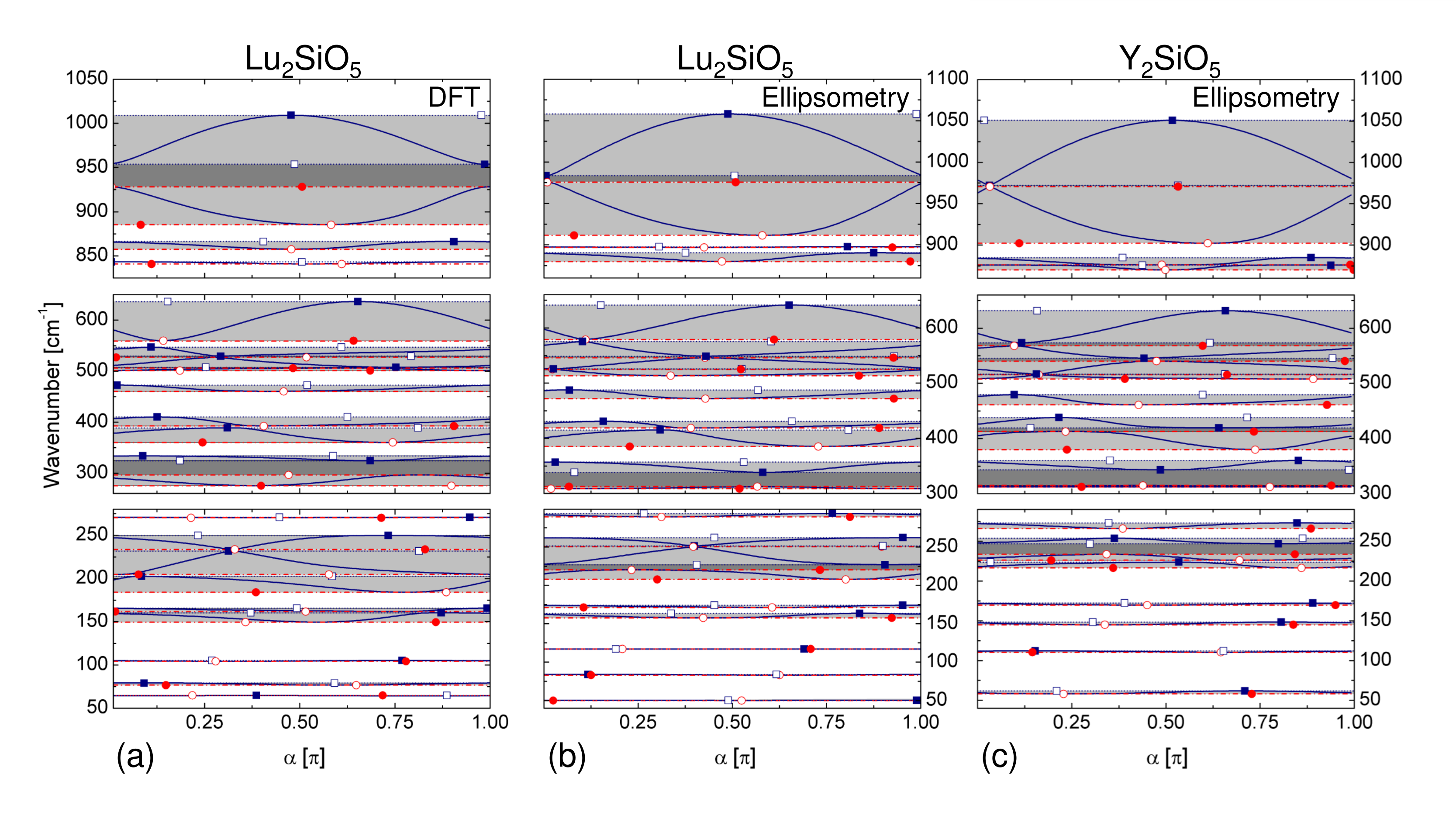}
\caption{\label{fig:orderLSO}Limiting infrared mode frequencies $\omega(\alpha)_{l}$ (blue solid lines) of monoclinic symmetry Lu$_2$SiO$_5$ as a function of unit direction
$\hat{\alpha} = \cos \alpha \hat{x} + \sin \alpha \hat{y}$ in the $\mathbf{a} -\mathbf{c}$ plane obtained from (a) density functional theory calculations, and (b) generalized ellipsometry investigations. Solid symbols (red circles: TO modes - also indicated by horizontal red dash-dot lines; blue squares: LO modes - also indicated by horizontal blue dotted lines) indicate frequencies ($\omega(\alpha_l)_{l}$) and eigenvector orientations ($\alpha_l$). Open symbols indicate the same but at directions normal to $\alpha_l$, i.e., at $\alpha_l \pm \pi$. Light gray areas indicate regions of so-called outer mode bands, dark gray areas indicate so-called inner phonon mode pairs. Outer mode bands cause polarized reststrahlen bands while inner mode bands cause unpolarized reflectance. See also Ref.~\onlinecite{SchubertPRB2019}. For comparison, (c) shows the ellipsometry investigation reported previously in Ref.~\onlinecite{Mock_2019} for Y$_2$SiO$_5$. We note overall excellent agreement between theory and experiment.}
\end{figure*}

In monoclinic symmetry materials with polar lattice vibrations the order of the phonon modes is directly related to the appearance of the polarized reststrahlen bands. The existence of inner and outer modes, where inner modes are nested within the frequency range of outer modes, and their relationship with the order of the phonon modes and the restrahlen range appearance was discussed recently for $\beta$-Ga$_2$O$_3$ as an example.\cite{SchubertPRB2019} The restrahlen bands for frequencies within outer modes are polarization-dependent. Inner modes cause polarization-independent (totally reflective regardless of polarization) reststrahlen bands. The directional limiting frequencies within the Born-Huang
approach are bound to within outer mode frequency regions not occupied by inner mode pairs. Early observations were reported by Kuzmenko for monoclinic copper monoxide and bismuth monoxide.\cite{KuzmenkoThesis2000} Hence, an unusual
phonon mode order can occur where both lower-frequency as well as upper-frequency limits for the directional modes can be both TO and/or LO modes. Figure~\ref{fig:orderLSO} depicts the directional mode frequencies and their dispersion within the monoclinic plane for LSO, obtained by DFT and by GSE, in comparison. Symbols indicate frequency and direction of all TO and LO modes (open symbols indicate their normal directions perpendicular to the eigenvectors within the monoclinic plane). Light gray areas indicate regions of outer modes not occupied by inner modes, and within which all directional modes are confined. The dark gray areas indicate regions occupied by inner modes, within which no directional mode exists. Overall, an excellent agreement between DFT and GSE is noted. Both order of modes, their frequencies as well as direction parameters are highly consistent between theory and experiment. Small deviations can be seen for small polarity modes which show very small dispersion only (outer or inner mode pairs with small TO-LO splitting). Also, some outer modes appear shifted and overlapping partially in the experiment, causing nested inner mode pairs and bands of total reflection (dark gray areas). 

We further compare the experimentally determined phonon mode properties of LSO with those of its isostructural compound YSO, whose phonon modes we have determined recently.\cite{Mock_2019} Yttrium (atomic number 39; electron configuration [Kr] 4d1 5s2; standard atomic weight 88.90584$u$; covalent radius $190 \pm 7$~pm) has nearly half of the inertial mass of Lutetium (71; [Xe] 4f14 5d1 6s2; 174.9668$u$; $187 \pm 8$~pm) but equivalent covalent size. As can been in Tab.~\ref{Table:ExpAu} in comparison with Tab.~4 in Ref.~\onlinecite{Mock_2019}, the A$_{\mathrm{u}}$ TO modes are very similar while the LO modes are slightly different reflecting slightly different Born effective charges. Figure~\ref{fig:orderLSO} also depicts the directional mode frequencies and their dispersion within the monoclinic plane for YSO, obtained by GSE. Comparing results from experiment for both compounds, one can see no significant differences between LSO and YSO, except for small variations in actual frequency and direction parameters. Closer inspection of Figs.~\ref{fig:TOphonons} and~\ref{fig:LOphonons} as well as Figs. 2 and 3 in Ref.~\onlinecite{Mock_2019} reveals that most of the phonon mode displacements in both LSO and YSO are taken up by the oxygen atoms, followed by much smaller displacement performed by the silicon atoms. The much heavier elements Y and Lu remain practically fixed within the lattice. Hence, replacement of Lu with Y affects the phonon mode behavior only marginally. A similar behavior was reported for Raman modes measured in LSO, YSO, and Lu$_{1.8}$Y$_{0.2}$SiO$_5$.\cite{Chiriu_2007} Therefore, we expect similar phonon mode behavior in all rare earth oxyorthosilicates Dy$_{2}$SiO$_{5}$, Ho$_{2}$SiO$_{5}$, Er$_{2}$SiO$_{5}$, Tm$_{2}$SiO$_{5}$, and Yb$_{2}$SiO$_{5}$ than observed here for LSO.

\subsubsection{Static and high-frequency dielectric constants}
\begin{table}
\caption{\label{tab:constants} Static and high-frequency dielectric constants obtained from DFT and GSE analyses reported in this work.}
\begin{ruledtabular}
\begin{tabular}{{l}{c}{c}{c}{c}}
    & $\varepsilon_{\mathrm{xx}}$ & $\varepsilon_{\mathrm{yy}}$ & $\varepsilon_{\mathrm{xy}}$ & $\varepsilon_{\mathrm{zz}}$ \\
  \hline
$\varepsilon_{\infty}$ & 3.16(6) & 3.12(7) & 0.002(7) & 3.23(9) \\
$\varepsilon_{\mathrm{DC}}$ & 12.02(9) & 10.65(8) & -0.85(1) & 13.65(3) \\
\hline
$\varepsilon_{\infty,\mathrm{DFT}}$ & 3.379 & 3.426 & -0.023 & 3.362 \\
$\varepsilon_{\mathrm{DC},\mathrm{DFT}}$ & 9.988 & 14.49 & 0.6685 & 12.21 \\
\end{tabular}
\end{ruledtabular}
\end{table}

Table~\ref{tab:constants} depicts results for the static and high-frequency dielectric constants obtained in this work. The static dielectric constants are obtained numerically by setting $\omega=0$. The high-frequency dielectric constants are determined during the best-match model GSE analysis. We note that the S-LST relation\cite{Schubert_2016_LST} is satisfied when considering all TO and LO mode frequencies and static and high-frequency dielectric constants obtained in this work.

\section{Conclusions}
We have determined the infrared active phonon mode parameters for the monoclinic symmetry rare-earth oxyorthosilicate Lu$_{2}$SiO$_{5}$. A combined analysis method using density functional theory and spectroscopic ellipsometry was used. Our previously described approach to extract phonon mode parameters from monoclinic symmetry and hence highly anisotropic materials has been demonstrated as a versatile technique. We found all phonon modes anticipated by symmetry and in excellent agreement between theory and experiment. We determined all directional modes and established the phonon mode order with the $\mathbf{a}-\mathbf{c}$ plane. We further observe that the phonon mode properties of Lu$_{2}$SiO$_{5}$ are very similar to its isostructural compound Y$_{2}$SiO$_{5}$, despite a much larger inertial mass of Y relative to Lu. This observation is explained by the large mass difference between oxygen and silicon relative to the Y and Lu atoms. We anticipate a very similar phonon mode behavior among the entire class of rare-earth monoclinic oxyorthosilicates.

\section{Acknowledgments} This work was supported in part by the National Science Foundation under award DMR 1808715, by Air Force Office of Scientific Research under award FA9550-18-1-0360, by the Nebraska Materials Research Science and Engineering Center under award DMR 1420645, the Swedish Research Council VR award No. 2016-00889, the Swedish Foundation for Strategic Research Grant Nos. FL12-0181, RIF14-055, EM16-0024, by the Knut and Alice Wallenbergs Foundation supported grant 'Wide-bandgap semi-conductors for next generation quantum components', and by the Swedish Government Strategic Research Area in Materials Science on Functional Materials at Link{\"o}ping University, Faculty Grant SFO Mat LiU No. 2009-00971. M.~S. acknowledges the University of Nebraska Foundation and the J.~A.~Woollam~Foundation for financial support. This research was performed while author A.M. held an NRC Research Associateship award at the U.S. Naval Research Laboratory. 


%

\end{document}